\documentstyle[amsfonts,preprint,aps]{revtex}

\begin{document}
\title{Numerical computation of resonance poles in scattering theory}
\author{Didier Felbacq}
\address{LASMEA UMR-CNRS 6602\\
Complexe des C\'{e}zeaux\\
63177 Aubi\`{e}re Cedex, France}
\maketitle

\begin{abstract}
We present a possible way of computing resonance poles and modes in
scattering theory. Numerical examples are given for the scattering of
electromagnetic waves by finite-size photonic crystals.
\end{abstract}

\bigskip \tightenlines

Resonance poles are the main quantity of interest in scattering theory \cite
{melrose,reed}. They are poles of the scattering matrix, and may be
considered as generalized eigenvalues to which generalized eigenmodes are
associated. These poles are complex ones, i.e. they correspond to complex
values of the energy in the scattering theory of Schr\"{o}dinger equation
and to complex frequencies for the scattering theory of Maxwell system.

When dealing for instance with finite size photonic crystals, i.e. periodic
arrangements of scatterers that are finite in at least one direction of
space \cite{joan}, one cannot use Bloch waves theory to compute the
electromagnetic properties of the structure and one has to retreat to the
computation of the scattering matrix, that is the operator $S\left( k\right) 
$ such that $U=S\left( k\right) U^{i}$ where $U^{i}$ is the harmonic
incident field, with wavevector $k$, and $U$ is the total field. The
scattering matrix writes $S\left( k\right) =I_{d}+T\left( k\right) $ where $%
T\left( k\right) $ is the so-called scattering amplitude. Let us assume that
there exists some pole $k_{p}$ of $T$ in some neighborhood ${\cal V}$ of the
complex plane, then locally the scattering amplitude writes $T\left(
k\right) =\frac{P_{p}}{k-k_{p}}+T_{0}\left( k\right) $ where $P_{p}$ is a
residu operator and $T_{0}$ is holomorphic in ${\cal V}$. Operator $P_{p}$
is a finite rank operator and its range is precisely the kernel of $%
T^{-1}\left( k_{p}\right) $. It is define in an abstract way as the Cauchy
integral: 
\begin{equation}
P_{p}=\frac{1}{2i\pi }\oint T\left( z\right) dz  \label{int1}
\end{equation}
where integration takes place on a loop oriented in the direct sense
enclosing the only pole $k_{p}$. Another way of defining the projection
operator $P_{p}$ is to define it as the following limit 
\begin{equation}
P_{p}=s-%
\mathrel{\mathop{\lim }\limits_{k\rightarrow k_{p}}}%
\left( k-k_{p}\right) T(k)
\end{equation}

The point of this note is to show that the first abstract definition (\ref
{int1}) can be turned into a useful numerical tool for both the computation
of the value of the pole and of the residu operator, and hence the
generalized eigenmode, whereas the second is useless. From a numerical point
of view, we of course only deal with finite rank operators and the
scattering amplitude admits a representation as an operator on $\ell
^{2}\left( {\Bbb Z}\right) $, that is as a matrix, in the usual meaning,
acting on double complex sequences\cite{moi,cent1}. Once this representation
is given, the residu operator can be computed provided that a region of the
complex plane containing only one pole can be precised. This means that it
suffices to know the value of the pole with a very poor precision to be able
to compute the residu operator, which is not the case when the second
definition (\ref{int2}) is used: in that last case numerical instabilities
necessarily occur as it uses the product of a singular matrix by a term
tending to zero, which is a very bad numerical situation. From a practical
point of view, one has to define a path $\gamma :t\in \left[ 0,1\right]
\rightarrow \gamma \left( t\right) \in {\Bbb C}$ whose graph is a loop
enclosing $k_{p}$ and to compute numerically the integral $%
\int_{0}^{1}T\left( \gamma \left( t\right) \right) \gamma ^{\prime }\left(
t\right) dt$ for which any reasonable numerical method works. However a
precise computation of the pole is useful when one wishes to compute a map
of the electromagnetic field of the pole, for in that case a particular
basis such as Hankel-Fourier series are used, i.e. the field is expanded on
the basis $\left( H_{n}^{\left( 1\right) }\left( k_{p}r\right) \exp \left(
in\theta \right) \right) _{n\in {\Bbb Z}}$ \cite{moi,cent1}. A possible way
is to use a M\"{u}ller like algorithm \cite{recipe}\ and to compute a zero
of the determinant of $T^{-1}\left( k\right) $. However this matrix is
generally badly conditionned and a better idea is to compute the smallest
eigenvalue of $T^{-1}\left( k\right) $. This works well in case of a finite
size crystal, but this is not always the case: for instance, when modelizing
photonic crystals by stacks of gratings \cite{mcphed} and introducing
periodic defects, convergence problems may occur when using M\"{u}ller
algorithm \cite{rafik}. We suggest then to compute the following Cauchy
integral: 
\begin{equation}
\frac{1}{2i\pi }\oint zT\left( z\right) dz=k_{p}P_{p}.  \label{int2}
\end{equation}
Recalling that $P_{p}$ has finite rank and hence has only a finite number of
eigenvalues, a simple comparison of this last integral with $P_{p}$ gives
the value of $k_{p}$ with a very good accuracy. Of course formula (\ref{int2}%
) only holds when $k_{p}$ is a pole with multiplicity $1$: this case is a
very common one. One should not mistaken the range of $P_{p}$ and the
multiplicity of $k_{p}$: it is possible that the multiplicity of $k_{p}$ is $%
1$ while the rank of $P_{p}$ is greater than $1$ \cite{felb}.

Let us now turn to some numerical applications. We deal with the structure
depicted in figure 1. It is a collection of $7\times 7$ homogeneous fibers
with relative permittivity $\varepsilon =9$, the radius of the rods $R=1/2$
and the spacing is $d=1$ (these values are given in arbitary units). We use
a rigorous modal theory of diffraction to compute the scattering matrix of
this system \cite{moi,cent1}. All the numerical results have been obtained
using a standard PC computer. Removing a rod at the center of the crystal, a
defect mode appears in the gap \cite{cent} (see fig. 2 for the transmission
spectrum). To this peak in the transmission spectrum corresponds a pole $%
k_{p}$. The reference value that we use for convergence comparison is $%
k_{p}=2.32919703586134-0.00378267987614i.$ This value has been computed
using M\"{u}ller algorithm by minimising the smallest eigenvalue of $T^{-1}$%
. For this given value of $k_{p}$ the smallest eigenvalue has modulus
inferior to $10^{-14}$. We then compute both Cauchy integrals (\ref{int1},%
\ref{int2}). We use an integration path that is a triangle whose vertices
have affixes: $(2.3,2.4-0.1i,2.4)$. We use the integration algorithm
described in \cite{numer} and we denote by $k_{N,p}$ the numerical value
obtained by using $N$ points of integration, which we compare with the above
value $k_{p}$ which is the best numerical value that we can obtain . A very
good precision is rapidly obtained (see fig. 3): for instance, using a
discretisation of $15$ points we obtain $6$ exact figures though with such a
rough discretization we only get operator $P_{p}$ with a low precision. In
fact, it seems that the proportional coefficient between both integrals is
not much affected by the precision with which $P_{p}$ is computed. A finer
computation of integral (\ref{int1}) gives the defect mode. The convergence
can be checked by looking at the non-zero eigenvalue of $P_{p}$ (fig. 4),
here the reference eigenvalue (i.e. the best numerical value for a precision
error of $10^{-15}$) is obtained with $N=150$. A much finer discretization
than in the case of the pole is required to get a good representation of the
defect mode, though the computing time is perfectly accessible with a very
basic PC.

In conclusion, we have shown that it was possible to turn a rather abstract
mathematical object into a useful numerical tool. This technique applies as
well for any situation in which a meromorphic operator with non essential
poles is involved, which is the usual case.

\newpage

{\large Figures captions}

Figure 1: Sketch of the 2D photonic crystal. The transmission ratio is
computed as the flux of the Poynting vector through the segment indicated
below the crystal.

Figure 2: Transmission ratio versus the wavenumber for an incident plane
wave.

Figure 3: Convergence of the value of the pole versus the number of
integration points.

Figure 4: Convergence of the eigenvalue of the projection operator versus
the number of integration points.

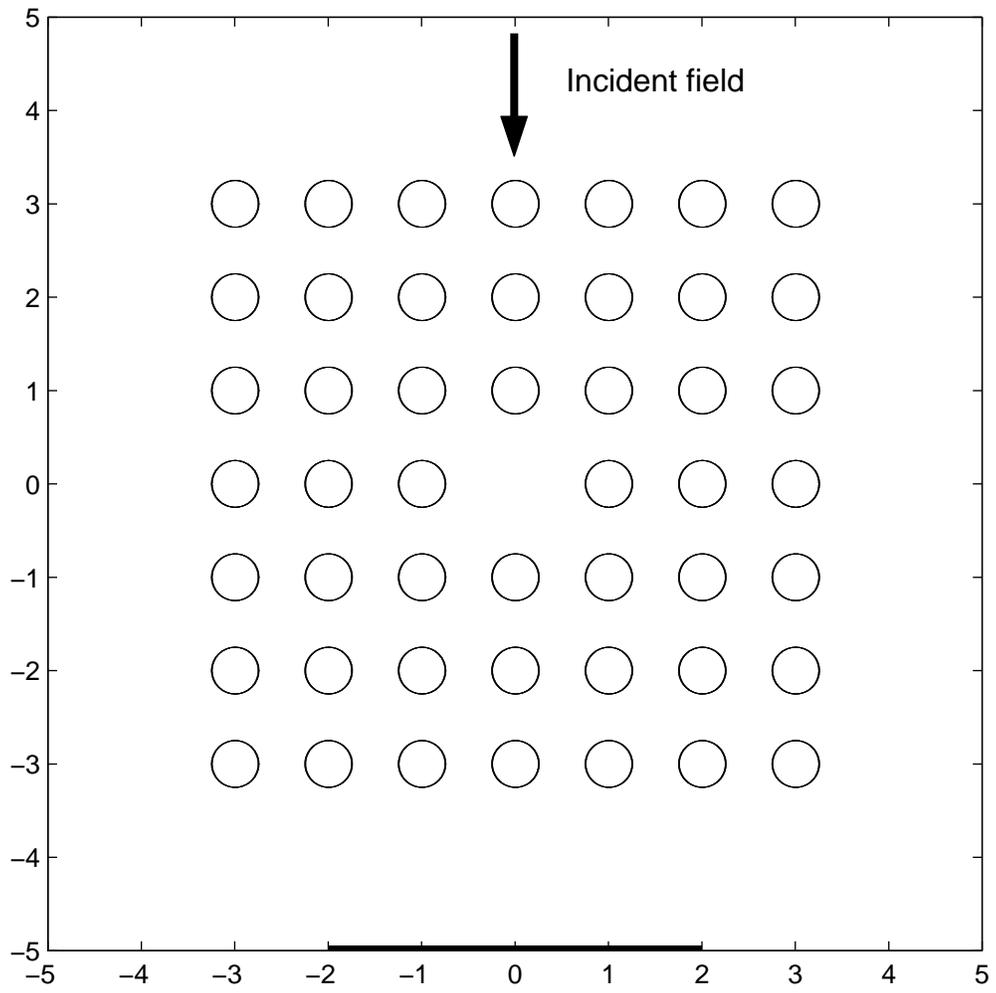
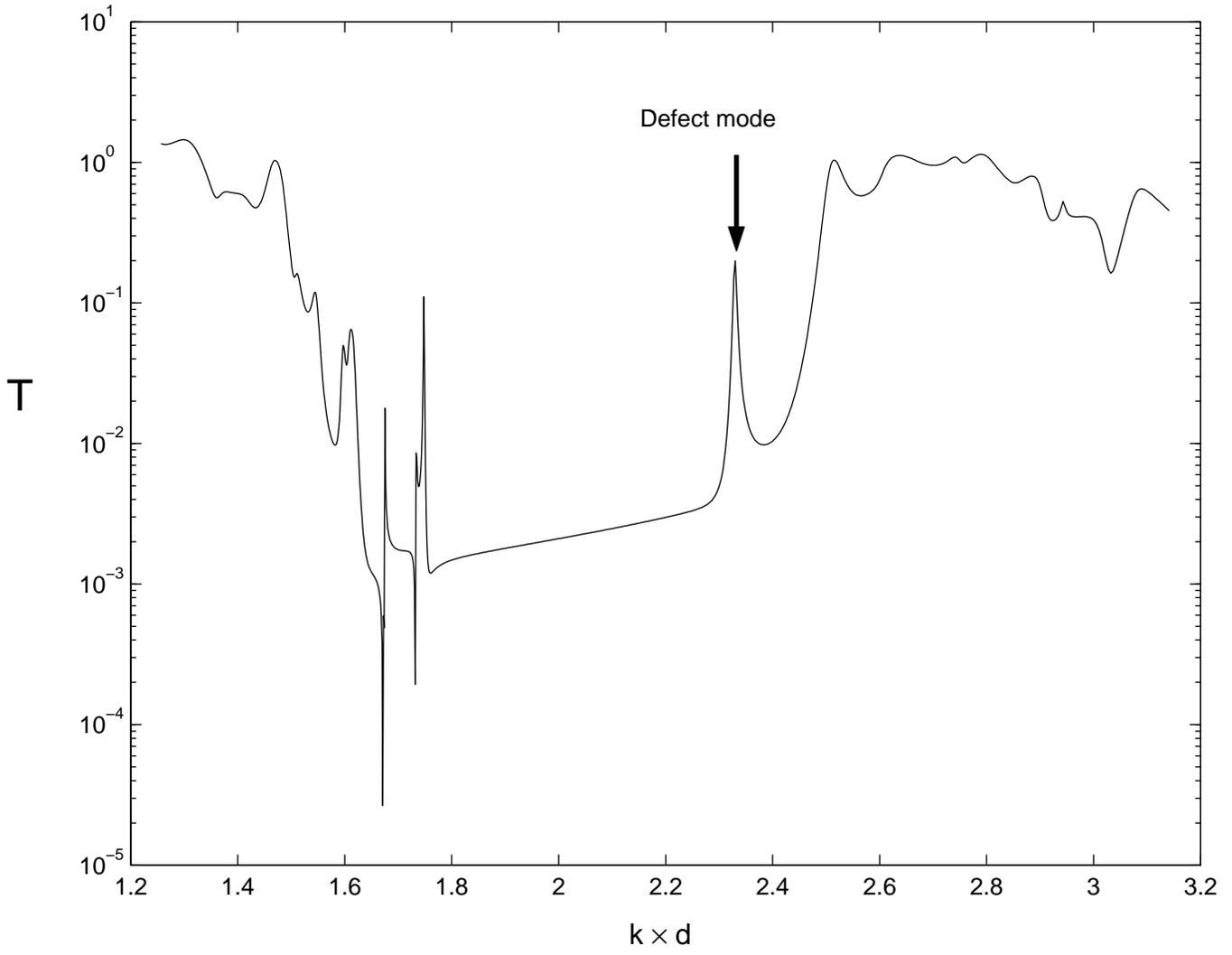
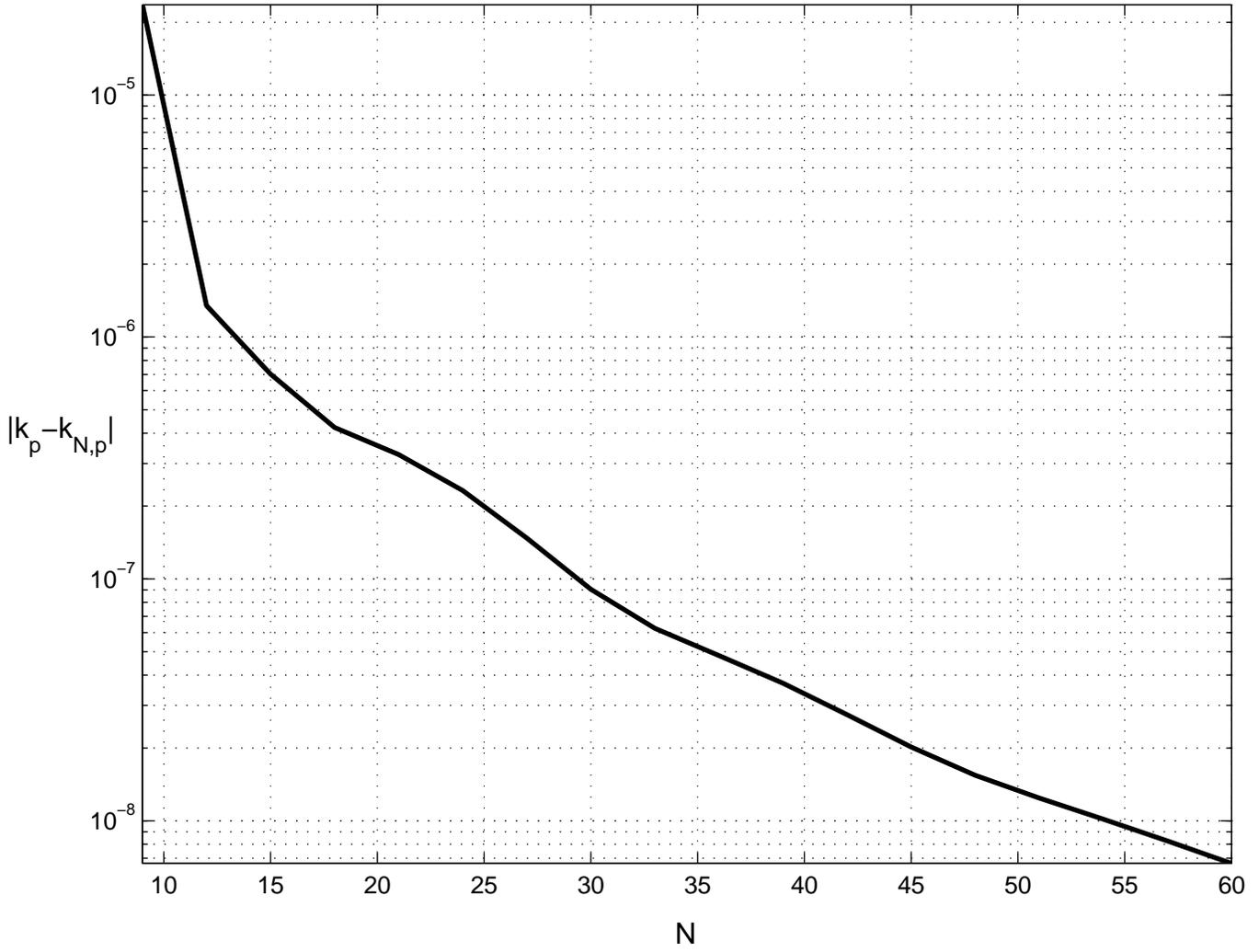
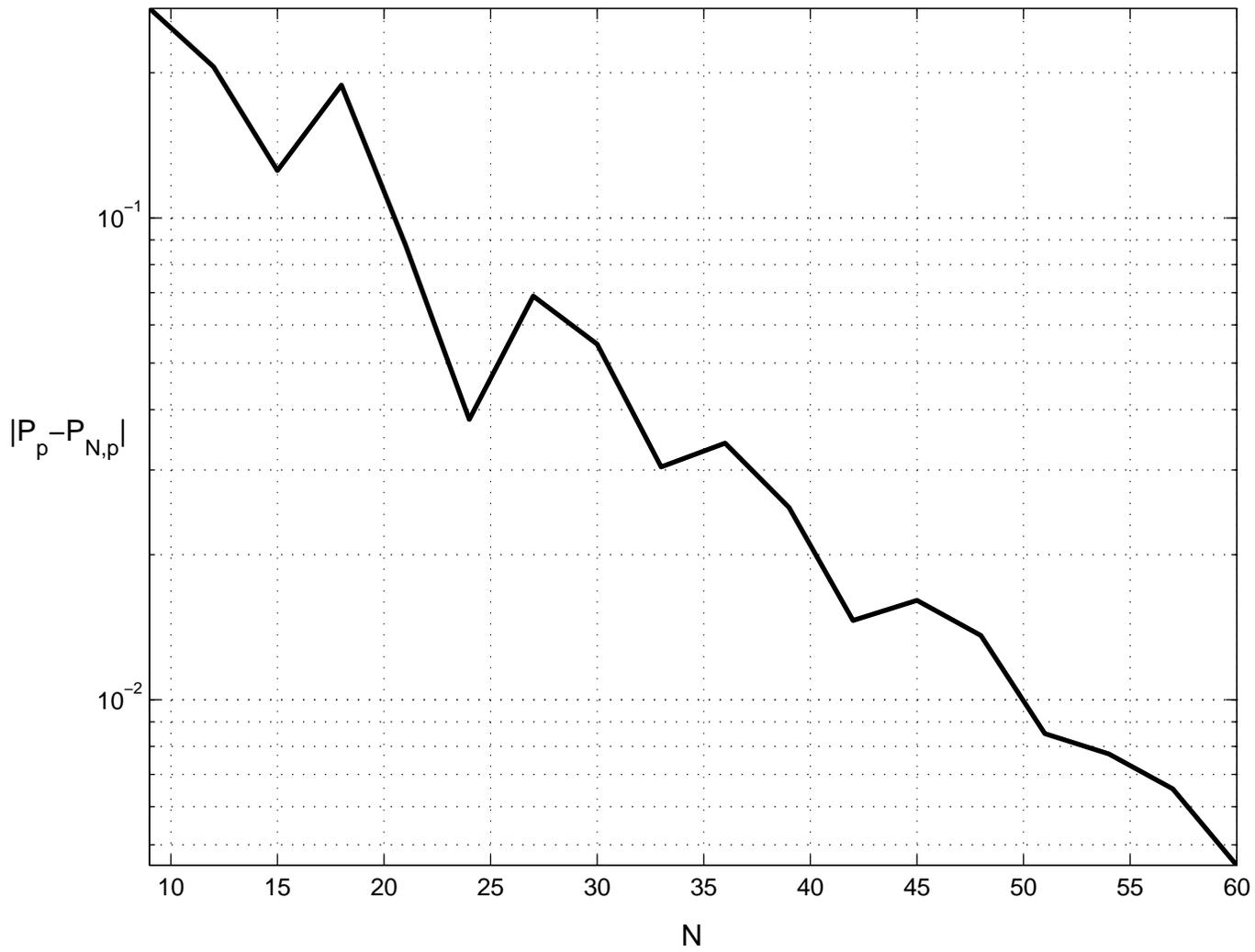
\end{document}